\documentclass[acmlarge,screen]{acmart}

\AtBeginDocument{%
 \providecommand\BibTeX{{%
  \normalfont B\kern-0.5em{\scshape i\kern-0.25em b}\kern-0.8em\TeX}}}

\acmPrice{0.00}
\acmISBN{}
\copyrightyear{2023} 
\acmYear{2023} 
\setcopyright{rightsretained} 
\acmConference[HCI4Climate Change]{Workshop at CHI'23}{April 28th, 2023}{Hamburg, Germany}
\acmBooktitle{HCI4Climate Change Workshop at CHI'23, April 28th, 2023, Hamburg, Germany} 
\acmDOI{}
\acmISBN{}

\begin{document}

\title{User Experience Considered Harmful (for the Planet)}


\author{Markus Löchtefeld}
\affiliation{%
 \institution{Aalborg University}
 \city{Aalborg}
 \country{Denmark}}
\email{mloc@create.aau.dk}

%
% Information:
% https://chi2023.acm.org/for-authors/alt-chi/
%
%
%
%
%

\renewcommand{\shortauthors}{Löchtefeld}

\begin{abstract} % less than 140 words
Great user experience is killing us (more or less)! My argument in this provocation is that the excessive focus on user experience (UX) by the tech industry and academic community has a negative impact on the sustainability of ICT devices. I will argue based on two examples, that we need new metrics or extend current UX metrics to also include third order effects and sustainability perspectives. Lastly, I would like us - the (Sustainable) HCI community - to increase our focus on solving the problems that result from our very own creations.
\end{abstract}



\maketitle

\section{Introduction}
\textit{In his 1968 influential letter "Go To Statement Considered Harmful"~\cite{dijkstra1968letters} Dijkstra criticised the - at that point ubiquitous - use of the Go To statement in programming languages and thereby laid the basics for structured programming. Ever after, papers with titles that include "considered harmful" have provoked and advocated for change around principles that are accepted and extensively used. The Human Computer Interaction (HCI) community also has seen several of such essays. For example Greenberg and Buxton argued that the use of usability evaluations not only can stifle innovation of novel user interfaces but also have potential detrimental effects on the scientific advances of HCI~\cite{greenberg2008usability}. In this provocation I do not aim to criticize the scientific process of user experience (UX) in HCI but rather - similar to Norman, who highlighted Activity-based Design as an alternative to Human-Centered Design~\cite{norman2005HCD} - question the detrimental effects that the obsession with great UX can have on the environmental impact of ICT products.}

The consequences of increased Green House Gas (GHG) emissions and the connected climate change have recently become much more apparent~\cite{ipcc2021climate}. While the Paris Agreement supposedly committed world leaders to ensure a maximum increase of 1.5\textdegree C of global warming, meeting this target seems increasingly less likely. The change needed to get close to this target will require drastic reductions across all sectors and ultimately reaching global net-zero GHG emissions by latest 2050. This will also mean that the Information and Communications Technology (ICT) sector will have to be decarbonized in some form or another. Freitag et al. examined peer-reviewed estimates on ICT's GHG emissions and their analysis suggest that they realistically equate to 2.1\% - 3.9\% of global GHG emissions~\cite{freitag2021real}. Furthermore, their analysis also highlights that without major efforts these emissions will not reduce, but are more likely to increase. ICT solutions are often suggested to help other sectors in reducing their GHG emissions~\cite{global2015gesi}, and several high-impact examples exist that demonstrate the potential, but whether the overall reductions will outweigh ICT's impact on GHG emissions is questionable~\cite{becker2015sustainability}. A particular negative example that is often brought forward is Bitcoin's energy- and e-waste-problems~\cite{de2022revisiting, de2021bitcoin}. This also leads Knowles et al. to point out that we should reconsider the digital exceptionalism treatment of ICT~\cite{knowles2022houseonfire}. Another larger problem that Freitag et al. mention, is that from ICT's overall emissions, ca. 30\% come from embodied emissions, meaning the GHG emissions released as part of the manufacturing process, e.g. extraction of raw materials, the manufacturing process and transport. Even if the other 70\%, emissions that stem from the use phase (energy consumption during use and maintenance) would be covered by renewable energy sources, the ICT sector could still be responsible for around 1\% of world-wide GHG emissions. 

The Sustainable HCI (SHCI) community has been for nearly 15 years~\cite{blevis2007sid, mankoff2007sigshci} actively engaged with mitigating issues that impact ecological- and social sustainability. In this provocation however, I will only focus on ecological sustainability, which however does not mean that I would argue for prioritizing one over the other. We have seen a lot of work focusing on ecological sustainability to reduce GHG emissions. Hannson et al. revealed that the Sustainable Development Goal (SDG) 12 "Responsible Production and Consumption" has been a major focus of attention~\cite{hansson2021decadeHCI} for example using persuasive technologies and eco-feedback~\cite{froehlich2010ecofeedback} to support users in saving energy~\cite{hagensby2021communalenergy}. And while this and many other examples exist that demonstrate the potential positive impact of the SHCI community, there has also been critique~\cite{bremer2022takenontoomuch,disalvo2010landscape, brynjarsdottir2012unpersuaded}. One particular point of criticism is that SHCI should collaborate more widely across fields and disciplines~\cite{silberman2014NextSteps, chen2016strategy} and include more community organisations~\cite{silberman2015information} or macrostructures (e.g. municipalities)~\cite{mollenbach2012macrostructures}. For example Silberman et al. argued that “the processes that give rise to the issues indexed by the term sustainability are larger in time, space, organizational scale, ontological diversity, and complexity than the scales and scopes addressed by traditional HCI design, evaluation, and fieldwork”~\cite{silberman2014NextSteps} which implies that the SHCI community needs to broaden its perspective. Bremer et al. analysed how the SHCI community has responded to such critiques and found that "SHCI has shifted from second-wave approaches towards those anchored in the third wave, and has found creative ways to capture the complexity of (un)sustainability and broaden the SHCI agenda"~\cite{bremer2022takenontoomuch}. However, this move to engage with sustainability in a broader manner also meant that the SHCI community quickly left "core" HCI territory and neglected efforts that lay at the heart of HCI. There are several first- and second wave HCI~\cite{bodker2006secondwave, bodker2015thirdwave} problems connected to the UX of ICT products, that have significant effects for example on the longevity and energy consumption of these products that have been so far neglected. While I acknowledge the SHCI community's effort and the success we have seen in third- (and fourth) wave HCI context, in this provocation I will argue that there is still a lot of work that needs to be done in our own backyard that can have significant positive effects on the sustainability of ICT. 

In this provocation I argue that we (the HCI community) need to focus more on our own creations and their direct impact. Specifically our obsession with UX as a measure could be indirectly one of the main contributors for an increase of ICT's GHG emissions. However, the goal of this provocation is not to criticize but rather, through two examples highlight potentials for future research directions that are at the core of HCI and that hold potential for more environmental sustainable ICT products.

\section{User Experience - the root of all evil?}
The term User Experience has been discussed and defined in many ways. The 2010 released ISO standard 9241-210:2010 "Ergonomics of Human-System Interaction—Part 210: Human-Centred Design for Interactive Systems" defines it as "A person’s perceptions and responses resulting from the use and/or anticipated use of a product, system or service." (in chapter 2 terms and definitions)~\cite{ISO2010UX}. Very similar even though maybe a bit more vague, the Nielsen Norman Group defines it as:"User experience encompasses all aspects of the end-user's interaction with the company, its services, and its products"\footnote{https://www.nngroup.com/articles/definition-user-experience/}. These definitions imply that UX designers need to adopt a holistic perspective on the design of all elements involved in the interaction, to create products that will be successful. 

In the SHCI community, Mankoff et al. proposed two ways to classify work, namely \textit{sustainability in design} - referring to approaches that focus on mitigating effects stemming from the material design of a product - and \textit{sustainability through design} - referring to approaches aim at creating more sustainable lifestyles empowered through technology~\cite{mankoff2007sigshci}. These classes are also sometimes referred to as \textit{sustainability by form} and \textit{sustainability by function}~\cite{Vissonova2018}. However, with respect to ICT products, such as smartphones, tablets, laptops or other smart-products, this differentiation can be problematic. As discussed above, UX adopts a holistic perspective and the UX design and evaluations of these products are usually a result of the combination of hardware and software. I argue that this artificial division of the SCHI community - while helpful for classifying novel third wave HCI approaches - can have detrimental effects on sustainability of ICT devices. As we use UX as a measuring stick for most of endeavours, we should use a similar holistic perspective on accessing the sustainability of the developed artefacts. 

While I agree with the fact that great UX is important for any product to succeed, there can arise problems from over-prioritizing it. Currently, UX is mostly measured in the first- and second order effect created by the use of the product~\cite{european2002impact}. First order effect means the impact and opportunities created by the existence of the product and second order refers to the ongoing use of the product. However, optimizing for short-term gains often can run counter to long-term success~\cite{neumann2012longterm}. As HCI researchers we often neglect third order effects, meaning the aggregated impact created by large number of users using the product over a long time~\cite{european2002impact}. The reason is that these are harder to predict and at this points we have few- or no tools for assessing them. These can however have massive detrimental consequences for the environmental impact of ICT products. Here we can see a first working point, UX evaluation methods that include a sustainability lens as well. I am not the first to argue for this, Thomas et al. already questioned the anthropocentric nature of the ISO standard 9241-210:2010~\cite{thomas2017hcdiso}. Remy at al. investigated the current limitations of evaluation possibilities~\cite{remy2017limitsofEval} and identified five key elements that can provide guidance to identifying evaluation methods~\cite{remy2018evalmethods}. Dourish framed UX in general as a legitimacy trap preventing HCI from its original goal of nurturing and sustaining human dignity and flourishing~\cite{dourish2019legitimacy}. While these suggestions are mainly meant to evaluate SHCI research, they would also help in all HCI and UX research work as well. And while integrating Life Cycle Assessments (LCA) as suggested in~\cite{remy2017limitsofEval,remy2018evalmethods} would help for example quantifying embodied GHG emissions and first- and second order effects quite well, LCA would be less suited for third order effects. Even Life Cycle Cost (LCC)- or whole-life cost analysis is quite hard to predict in cases of exponential user growth. At this point very few practical guidelines exist and the closest we have to applicable guidelines are from Dillahunt et al.~\cite{dillahunt2010proposed}, which don't directly consider third order effects either. So to this extend we need to work on more holistic evaluations metrics that allow to access UX and environmental sustainability with respect to third order effects. One particular example in user interface (UI) design, is if excess data is used to download something in the background that is never seen or interacted with (and not needed) on the user side. Considering exponential user growth this can in third order effects lead to quite drastic excess data and thereby excess GHG emissions.

%What does this mean for the sustainability of ICT devices? What do we have to do to decrease the GHG of ICT devices? Multiple pathways are possible if we ignore energy efficiency increases through new hardware. On the one hand we can decrease the energy consumption from the use phase through software (e.g., reduction of data traffic~\cite{frick2016designing} or creating more energy efficient software by using low level programming languages~\cite{pereira2017programming}) and on the other hand we can reduce the impact of embodied emissions by extending the longevity of the hardware (so that per use hour less new hardware is needed~\cite{becker2014sustainability}). 

\subsection{Hidden Data Traffic in UIs}
Hidden data traffic - so a system downloading data without the user explicitly telling it to do so - occur in many scenarios. Most of these can be considered a direct result of increasing the UX of a specific UI. Probably the most common one is the endless scrolling feature on social media sites like Facebook, Instagram or TikTok, where, before the user scrolls to the bottom, more posts are pre-loaded so that no visible loading time occurs. This endless scrolling dark pattern~\cite{gray2018darkpatterns,Monge2022darkpatterns} is often considered to lead to mindless scrolling stealing the users time, however, I don't want to engage with the ethical dimension of this but rather focus on the environmental impact. One might argue that in most cases their is only little data overhead of a few MB, which is most likely the case, however, given that there are between 2-3 billion monthly active social media users, this can result in a large amount of wasted data. An example: If we assume that for every daily active Instagram user (1.386 billion\footnote{https://blog.hootsuite.com/instagram-statistics}) one image is loaded in excess (in the authors short test using the web client of Instagram, on average 3 images were loaded that were not visible and the image sizes varied between 31kb and 475kb with an average of 173kb over 50 images) this would result in 0.239 petabyte of excess data. At this point I abstain from calculating energy consumption or GHG emissions, as the numbers vary so widely in literature~\cite{freitag2021real}. However, I hope that this estimate of 0.239 petabyte just for Instagram, demonstrate that the problem should not be ignored. Here we can see a third-order effect, turning a tiny negative element into a massive problem. Widdicks et al. highlighted a similar issue in their analysis of streaming services~\cite{widdicks2019streaming}, where HCI promotes excellence in UX, SHCI would demand more conscious utilisation of resources. In their work they discuss for example the reduction of video quality to reduce data usage when streaming, or for example including the possibility of not streaming the video, if only audio is wished for. 

A way to reduce such hidden data traffic could be to introduce design friction - elements that are less effortless and more taxing for the user to carry out~\cite{cox2016friction, benford2012uncomfortable}. Before the proliferation of 3G, most mobile social media apps, did not automatically download more content in the background and the user had to press a button to get more content. Such simple frictions would save potentially a lot of data traffic~\cite{frick2016designing} and thereby a lot of energy. However, there are multiple problems at this point. First we don't have a good understanding of the theoretical workings of design frictions, but it recently gained some more attention~\cite{gould2021frictionSIG}. Second, such frictions would be countering the business models of big social media companies. While the first issue can be overcome through more foundational research, the second probably won't. When simply using the websites of the social media companies, browser plugins might allow to add this friction and generate more awareness. Furthermore, there are many more cases (not only the endless scrolling) where data is pre-loaded for potential cases that don't happen, all in the name of UX.

This is a good example of a second-/third wave problem that would benefit from more thorough investigation by the (S)HCI community. Many cases of hidden data usage for the sake of UX exist that potentially could be solved by friction. While there are most likely cases with bigger potential gains, this example highlights a missing assessment of UX and potential unsustainable third order effects.

%At this point I don't want to even mention potential issues - in a social sustainability sense - that arise from not incorporating non-normative and diverse user groups, as this would be a whole paper by itself~\cite{spiel2018fitness}.

% furthermore software experiences becoming more "frictionless" e.g. in DOS days one could create specific boot disks, to load or not load drivers for hardware to use less memory

\subsection{Repair \& Longevity}

% tight integration -> e.g. smartphone becoming thinner and more integrated and less serviceable -> repairable smartphones are usually etwas bigger, Fairphone
The second example I want to discuss here is the effect that UX has on embodied emissions of the ICT sector. Our current linear way of producing, consuming, and discarding ICT products is not sustainable. To overcome this, several different pathways have been suggested in the past, for example, Slow Tech. This promotes a sustainable, ethical way of technology production and consumption~\cite{patrignani2014slow, patrignani2018slow} achievable by slowing down production rate, consumption, and disposal pace~\cite{lebel2016fast}. One particular important element to reduce the disposal pace is repair. This has also been discussed as one of the key sub strategies for sustainable production systems in the circular economy (CE)~\cite{macarthur2013towards}. SHCI has been focusing on repair as one of the key strategies to sustainability from the very beginning~\cite{blevis2007sid, huang2008repair} and multiple approaches have been explored~\cite{remy2015addressing, jackson2014artofrepair, ozccelik2022long}. Generally users are replacing their devices to early compared to their optimal lifetime~\cite{magnier2022replaced} and Brusselaers et al. also found that most users do not repair their broken products, even when it is possible and economically beneficial~\cite{brusselaers2020economic}. Prior research (particular stemming from industrial- and product design) shows that a user's decision is whether to repair a product or not is depending on the one hand on whether the product is designed for repair and on the other hand it relies on the user’s knowledge and skills~\cite{cooper2018fix, ackermann2018design, jaeger2021users}. Especially the high level of integration in current ICT devices is quite problematic, as it obfuscates the inner layers and makes it hard to identify faults in the product~\cite{arcos2021faults}. 

While the tight integration which e.g. decreases size and weight, and the usage of more energy demanding materials such as aluminium and glass, are beneficial for the hedonic qualities, aesthetics and feel of ICT products they are detrimental to their repairability. For example, a phone with a glass back can "feel" sleeker and more desirable, however a glass back usually requires some kind of glue instead of screws to be hold in place and also is much more fragile. However, solving these aspects should be left to industrial- and product design experts. 

There are however elements that the SHCI community can focus on and here I will postulate the hypothesis that if users have a structural mental model~\cite{westbrook2006mental} of an ICT product, they will be able to make better repair decisions and prolong the products lifetime. Arcos et al. already highlight the importance of the design in making repair decisions~\cite{arcos2021faults}. However, if we at ICT technology such as Smartphones and Laptops (here as the prime examples), we can see that these devices don't require the user to understand how they work internally, for them to be used. 20 years ago this was very different, users had to e.g., install operating systems and wrangle with hardware drivers, thereby gaining a deeper understanding of the inner workings. This is similar to current users that build their own (gaming) PC's. Their level of knowledge of how an ICT device works is usually much higher, so they are much more likely to try to repair issues and replace components~\cite{arcos2021faults}. Following this argument, we see that good UX - here reducing the friction and making the technology easily accessible for non-experts - is advancing the technologies unsustainability. Obviously, I will not argue that we should make the entry into using a computer harder again, but it would be worthwhile to understand how UI design and timed nudges during usage could create a structural mental model that would allow users to make better repair choices. Similar to the previous example, to support repair activities devices might also require some more friction in the interface to make users reflect on the inner workings of the devices they are using~\cite{gould2021frictionSIG}. Devices like the Fairphone\footnote{https://www.fairphone.com/} or the Framework Laptop\footnote{https://frame.work/} are designed to be repairable and upgradeable, it is part of the brand and communicated in a way that it will part of the mental model the customer will get when they purchase the product. Framework even has a DIY edition, where the user has to assemble some of the parts before being able to use it\footnote{https://frame.work/products/laptop-diy-12-gen-intel}. This means that the user also immediately gets an understanding of what is possible to repair or exchange. These are the kind of interactions that I envision could be worthwhile on a hardware level. On an Operating System level, Linux distributions (while having worked very hard to increase the UX and reduce friction) could be modified to allow users to get a better understanding of, and enable deeper interaction with, the inner workings of the computer for different experimental settings. 

The case of repair is another first/second wave HCI problem where the SHCI community can have a significant impact. While it is a much more complex problem with a variety of different layers that will require huge efforts to properly investigate and solve the issue, it is a very promising area, with potentially significant impact.

\section{Conclusion}
My aim with this provocation is to highlight that before taking on more, we as the (S)HCI community should focus on the effects of our own doing first. My answer to Bremer et al.~\cite{bremer2022takenontoomuch} would be, yes the community is taking on too much. It might be what Knowles et al. call digital exceptionalism treatment of ICT~\cite{knowles2022houseonfire}, that drives the community to solve a lot of other problems before we focus on ICT itself. An important first move, should be to establish, inside the HCI community, that UX as the ultimate metric will harm sustainability of ICT devices, and that we need to take on more extensive and further reaching metrics. Designers must recognize that humans do not typically occupy a central role, but rather are part of a complex network of human and non-human actors that are socially, economically, and ecologically interdependent~\cite{coulton2019more}. And this needs to be also reflected in our tools to design and analyse interfaces.

What do we have to do to decrease the GHG of ICT devices? Multiple pathways are possible if we ignore energy efficiency increases through new hardware. On the one hand we can decrease the energy consumption from the use phase through software (e.g., reduction of data traffic~\cite{frick2016designing} or creating more energy efficient software by using low level programming languages~\cite{pereira2017programming}) and on the other hand we can reduce the impact of embodied emissions by extending the longevity of the hardware (so that per use hour less new hardware is needed~\cite{becker2014sustainability}).

\textit{I do not want this provocation to be understood as critique of the (S)HCI or UX community. Engagement with outside communities and experts is essential for enabling sustainable lifestyles that we desperately need in the future. However, I would like to remind (S)HCI researchers, not to forget the problems that result from their own creations.}

\bibliographystyle{ACM-Reference-Format}
\bibliography{references}

\end{document}